\documentclass[runningheads]{llncs}
\usepackage{graphicx}
\usepackage{todonotes}
\usepackage{subcaption}
\usepackage{caption}
\usepackage[T1]{fontenc}

\begin{document}
\title{Citizen Science and Machine Learning for Research and Nature Conservation: The Case of Eurasian Lynx, Free-ranging Rodents and Insects}
\titlerunning{Citizen Science and Machine Learning for Nature Conservation Panel}
%

\author{
Kinga Skorupska\inst{1}\orcidID{0000-0002-9005-0348} \and
Rafał Stryjek\inst{2}\orcidID{0000-0003-3889-9341} \and
Izabela Wierzbowska\inst{3}\orcidID{0000-0001-6329-7241} \and
Piotr Bebas\inst{5}\orcidID{0000-0003-2956-5892} \and
Maciej Grzeszczuk\inst{1}\orcidID{0000-0002-9840-3398} \and
Piotr Gago \inst{1}\orcidID{0000-0001-7288-4210} \and
Jarosław Kowalski \inst{4}\orcidID{0000-0002-1127-2832} \and
Maciej Krzywicki\inst{6}\orcidID{0000-0002-8464-2830} \and
Jagoda Lazarek\inst{1}\orcidID{0009-0006-8747-2116} \and
Wiesław Kopeć \inst{1}\orcidID{0000-0001-9132-4171}
}

\authorrunning{Skorupska et al.}
%
\institute{XR Center, Polish-Japanese Academy of Information Technology
 \url{https://xrc.pja.edu.pl}
 \and Institute of Psychology Polish Academy of Sciences
\and Institute of Environmental Sciences, Faculty of Biology, Jagiellonian University
\and National Information Processing Institute
\and Faculty of Biology, University of Warsaw
\and Geografia Malownicza}
\maketitle              
\begin{abstract}

Technology is increasingly used in Nature Reserves and National Parks around the world to support conservation efforts. Endangered species, such as the Eurasian Lynx (\textit{Lynx lynx}), are monitored by a network of automatic photo traps. Yet, this method produces vast amounts of data, which needs to be prepared, analyzed and interpreted. Therefore, researchers working in this area increasingly need support to process this incoming information. One opportunity is to seek support from volunteer Citizen Scientists who can help label the data, however, it is challenging to retain their interest. Another way is to automate the process with image recognition using convolutional neural networks. During the panel, we will discuss considerations related to nature research and conservation as well as opportunities for the use of Citizen Science and Machine Learning to expedite the process of data preparation, labelling and analysis. 


\end{abstract}
\section{Rationale}

Technology is increasingly used in Labs, Nature Reserves and National Parks around the world to support research and conservation efforts. Endangered species, such as the Eurasian Lynx, are monitored by a network of automatic photo traps. This allows scientists to gather necessary data to observe migration patterns, evaluate sizes of populations and their distribution across vast areas \cite{norouzzadeh2018}. Yet, before this happens, scientists are left with large amounts of data, which needs to be prepared, analyzed and interpreted. Therefore, researchers working in this area increasingly need support to process this incoming information. One opportunity is to seek support from volunteer Citizen Scientists who can help label the data, however, it is challenging to retain their interest. Another way is to automate the process with image recognition using convolutional neural networks (CNNs). During the panel, we will discuss considerations related to nature conservation and opportunities for the use of Citizen Science and Machine Learning to expedite the process of data preparation, labeling, and analysis. Networks trained on image datasets that are already available can add preliminary labels, which later can be verified by citizen science volunteers \cite{drakshayini2023}, feeding this data back to further fine-tune the model as part of human-in-the-loop workflows. CNNs trained in this way can recognize not only images that contain animals, but also the specific species and number of animals, the habitat type, and even identify individual animals within species \cite{drakshayini2023}.

\section{Theme}

The key considerations of this discussion panel are related to ways of supporting scientists engaged in conservation and research efforts.

\subsection{Discussion Points}
\paragraph{Aspect 1: What Data do Scientists Need? Challenges of analyzing data from photo traps and video surveilance. }

The technological advancement has provided researchers with access to various types of inexpensive, reliable, and user-friendly video recording equipment, such as camera traps and video surveillance, which by nature do not discern the types of events they capture. They are simple devices equipped with movement sensors (see Figure 3) that may be triggered by a stronger gust of wind moving surrounding leaves, an insect or changes in solar illumination related to the movement of clouds. That means that they capture many more images and videos than necessary - many of them are seemingly empty, but to determine this, scientists still have to view them in search of the animals they wanted to track. But their task does not stop there, once they find an image or a video with an animal, they want to know the species, age, gender, how many there are, what they are doing, are there many different species present and what are the circumstances: day, night, snow, rain.

\paragraph{Aspect 2: What does the workflow look like now? The potential to involve Citizen Scientists.}

Currently, a lot of photos and videos taken by photo traps or recorded by surveillance systems are analyzed by scientists. The use of citizen science volunteers to facilitate this process could free up scientists' resources to focus on tasks requiring expert knowledge. However, citizen science task design ought to take into account the needs of the contributors, to retain their engagement.

\paragraph{Aspect 3: Big Data and Automation: Using Machine Learning in Nature Conservation}
Machine learning is a great way to complement image identification and video analysis workflows involving humans. In general, with ML it is possible to evaluate whether an image or a video contains an animal, and often, what species it is. For Eurasian Lynx, the models pre-trained on tagged domestic cat (\textit{Felis catus}) photos had quite high accuracy, barring extreme cases, such as difficult weather conditions and bad light. However, the task that is very challenging to ML still remains, that is the identification of individual animals. Yet, at the same time, this is the task that researchers are the most interested in, to study the behavioral patterns of these individuals so that conservation efforts can be realized better.

   \begin{figure}[!ht]
    \setkeys{Gin}{width=0.31\linewidth}
    \captionsetup[subfigure]{skip=0.5ex,
                             belowskip=1ex,
                             labelformat=simple}
    \renewcommand\thesubfigure{}
    \newcommand\sep{\hspace{0.025\linewidth}}

    \subfloat[Figure 1: Image taken with a Sony A58 camera with Tamron SP 70-300 mm f/4-5,6 Di USD Sony lens in good light conditions where it is easy to distinguish a pair of Roe Deer. Photo by Maciej Krzywicki reproduced with permission.]{\includegraphics{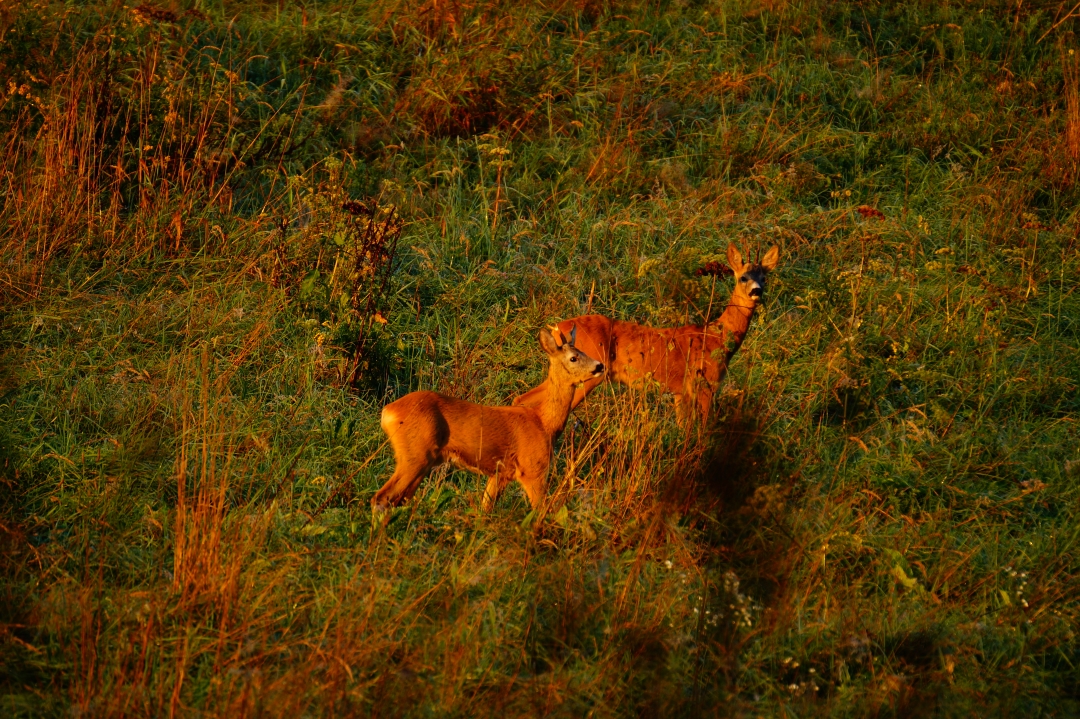}}
    \sep
    \subfloat[Figure 2: Image taken with a TopHunt HC900A 1080p 36MP photo trap with three Red Deers present, of which only two are easily distinguishible. Still of a photo trap video by Maciej Krzywicki reproduced with permission.]{\includegraphics{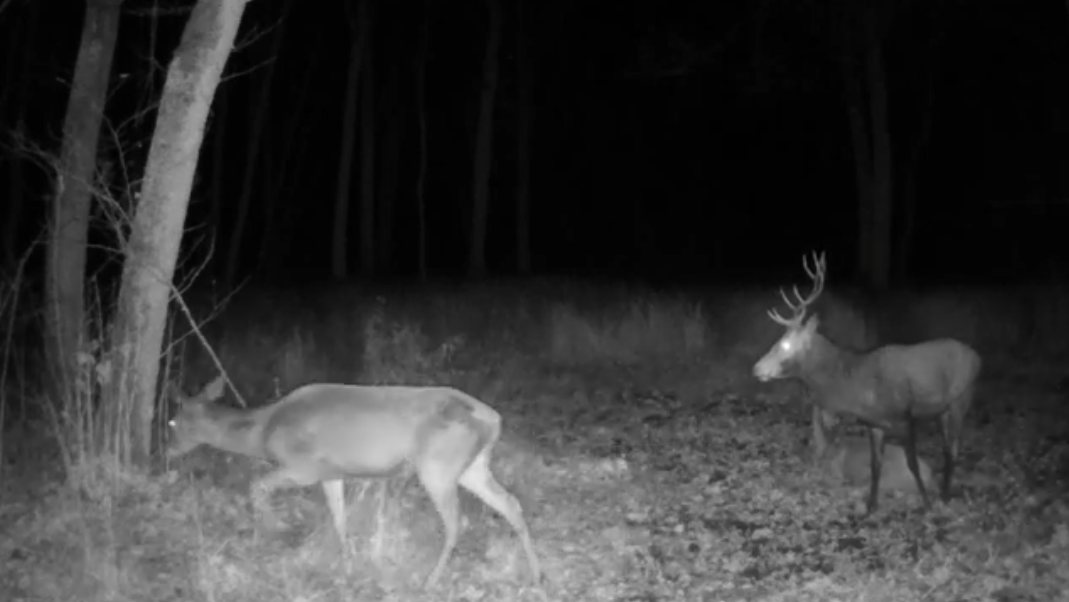}}
    \sep
    \subfloat[Figure 3: A photo trap with its architecture where LED is an infrared lamp, L the lens and PIR a movement sensor. Image by Dariusz Kowalski, licensed under CC BY-SA 4.0]{\includegraphics{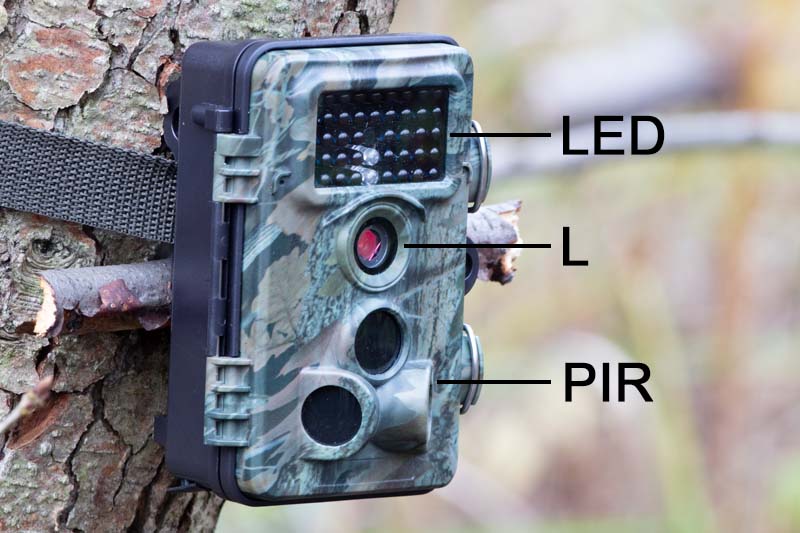}}

    \end{figure}

\begin{figure}[!ht]
    \setkeys{Gin}{width=1\linewidth}
    \captionsetup[subfigure]{skip=0.5ex,
                             belowskip=1ex,
                             labelformat=simple}
    \renewcommand\thesubfigure{}
    \newcommand\sep{\hspace{0.025\linewidth}}

    \subfloat[Figure 4: An example of a study where video analysis using machine learning would greatly benefit researchers, primarily in terms of species identification and measurement of the time spent exploring the experimental area. Additionally, it would aid in identifying individual animals and marking their individual behaviors, and movement tracking. A) Two wooden chambers deployed next to a forest and meadows. (B) A CCTV camera recording the entrance area to the chambers. (C) An interior of one of the chambers with an entrance pipe, bait, and a CCTV camera. (D) A striped field mouse (\textit{Apodemus agrarius}). (E) A yellow-necked mouse (\textit{Apodemus flavicollis}). (F) A bank vole (\textit{Clethrionomys glareolus}). (G) A video frame showing two striped field mice feeding inside a chamber. (H) A video frame showing two yellow-necked mice feeding inside a chamber. (I) A video frame showing two bank voles inside a chamber. Video frames taken by infrared cameras with motion detection (Easycam EC-116-SCH; Naples, FL, United States) connected to digital video recorder (Easycam EC-7804T; Naples, FL, United States). Photos by Rafał Stryjek.]{\includegraphics{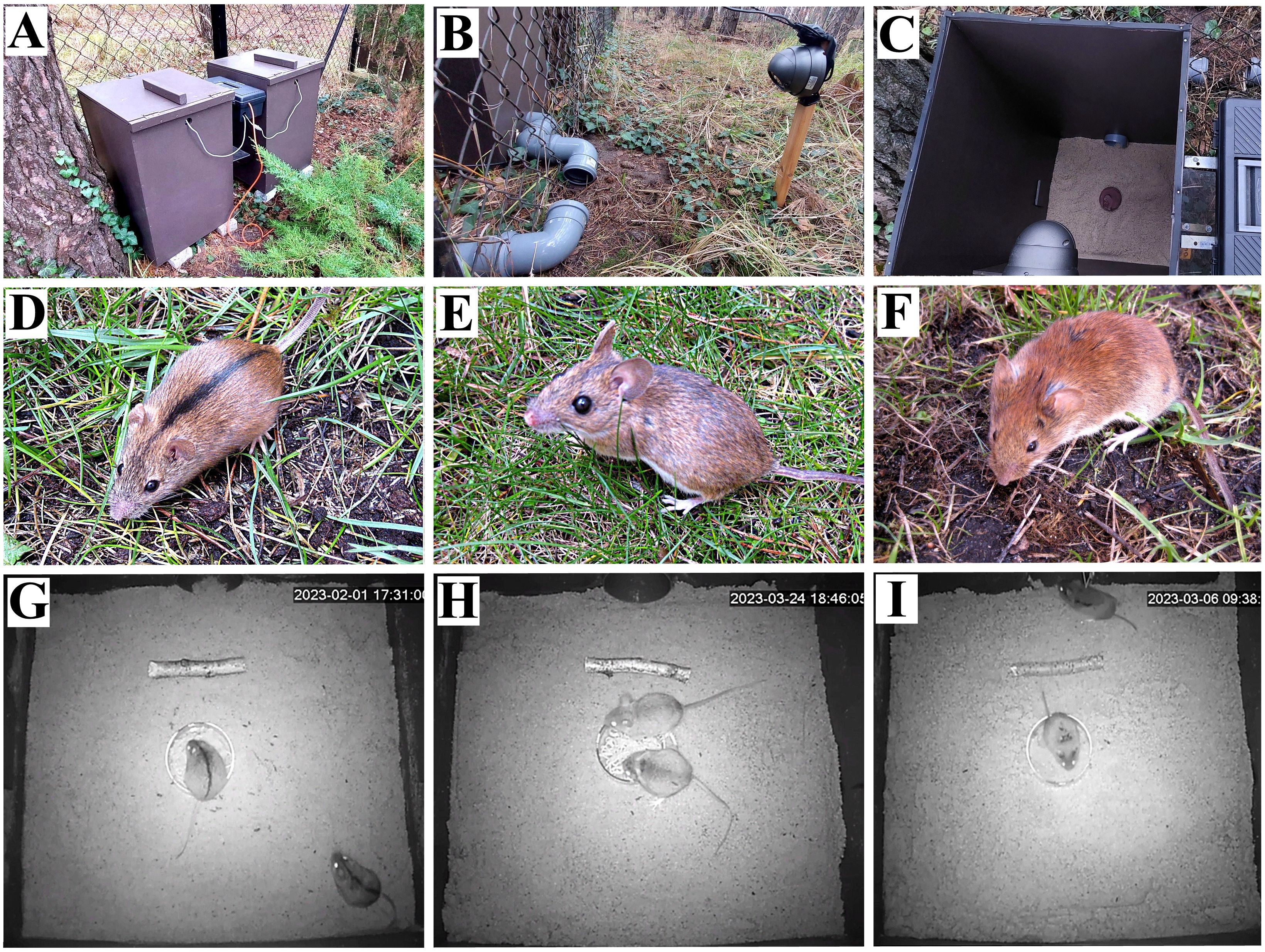}}
    \end{figure}

\begin{figure}[!ht]
    \setkeys{Gin}{width=1\linewidth}
    \captionsetup[subfigure]{skip=0.5ex,
                             belowskip=1ex,
                             labelformat=simple}
    \renewcommand\thesubfigure{}
    \newcommand\sep{\hspace{0.025\linewidth}}

    \subfloat[Figure 5: Exemplary video frames and photos of butterflies, which, as per the experimental protocol, should be differentiated as four distinct individuals belonging to two different species, would greatly benefit from machine learning-supported recognition for such comparisons. A) \textit{Iphiclides podalirius} - likely an individual shortly after eclosion captured using the LIVE-cam option of the iPhone 13 Pro, featuring clearly visible hindwing tails and wing edges; B) \textit{I. podalirius} individual captured with the same camera on the same day but probably older and after experiences, possibly following a predator attack, with the absence of both hindwing tails and a stripped lower edge of the right hindwing; C) \textit{Gonepteryx rhamni} – a male photographed with the HTC Desire 21 Pro 5G phone camera on the scarlet runner bean (\textit{Phaseolus coccineus}) flower, displaying distinct wing venation and brown-orange spots that may vary in the color intensity and size, allowing for individual distinctions; D) \textit{G. rhamni} – another male photographed on the heliotrope (\textit{Heliotropium arborescens}) flower arranged and illuminated to make wing venation and spots less noticeable; machine learning-supported recognition would excellently distinguish both males, such as by identifying differences in the position of spots on the wings in relation to other body parts. Photos by Piotr Bębas.]{\includegraphics{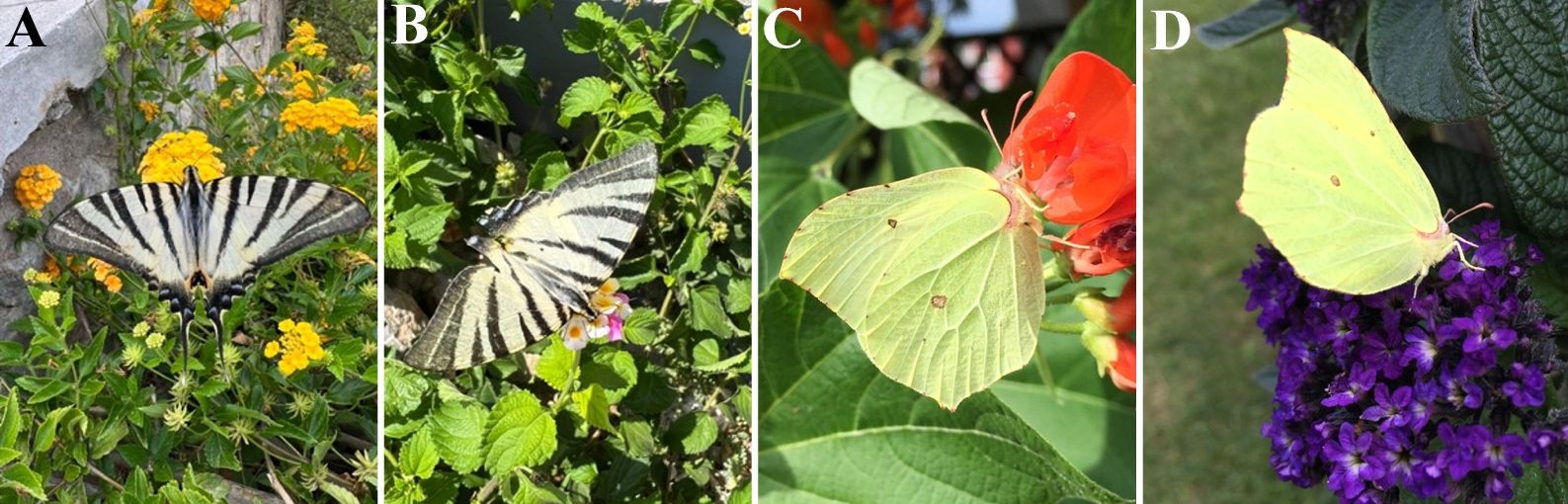}}
    \end{figure}

\begin{figure}[!ht]
    \setkeys{Gin}{width=1\linewidth}
    \captionsetup[subfigure]{skip=0.5ex,
                             belowskip=1ex,
                             labelformat=simple}
    \renewcommand\thesubfigure{}
    \newcommand\sep{\hspace{0.025\linewidth}}

    \subfloat[Figure 6: iPhone 13 Pro photos and video frames of butterflies representing five different species, relatively similar in coloration and wing patterns, which regularly visit flower beds serving as experimental fields – for their recognition and assignment to specific species, a tool based on machine learning-supported recognition will be essential. A) \textit{Pararge aegeria} with a distinctly orange pattern on a brown background and with moderately scalloped (sinusoidal) wing margins; B) \textit{Polygonia c-album} with orange wings featuring a brown pattern and sharply scalloped edges; C) \textit{Argynnis paphia}, also with orange wings and a brown pattern, with colors similar in shade to those of the other shown species but with clearly the least scalloped wing edges; D) LIVE-cam video frame of \textit{Colias croceus} representing a species with distinct eyespots and characterized by significant color and size/shape variability of eyespots; E) Video frame of \textit{Maniola jurtina} with similar characteristics to the latter in terms of coloration and eyespot pattern (position and morphology), which, with complex wings (as shown in the photos), may pose a challenge for automatic species recognition, a challenge that an advanced system based on machine learning-supported recognition can solve. Photos by Piotr Bębas.]{\includegraphics{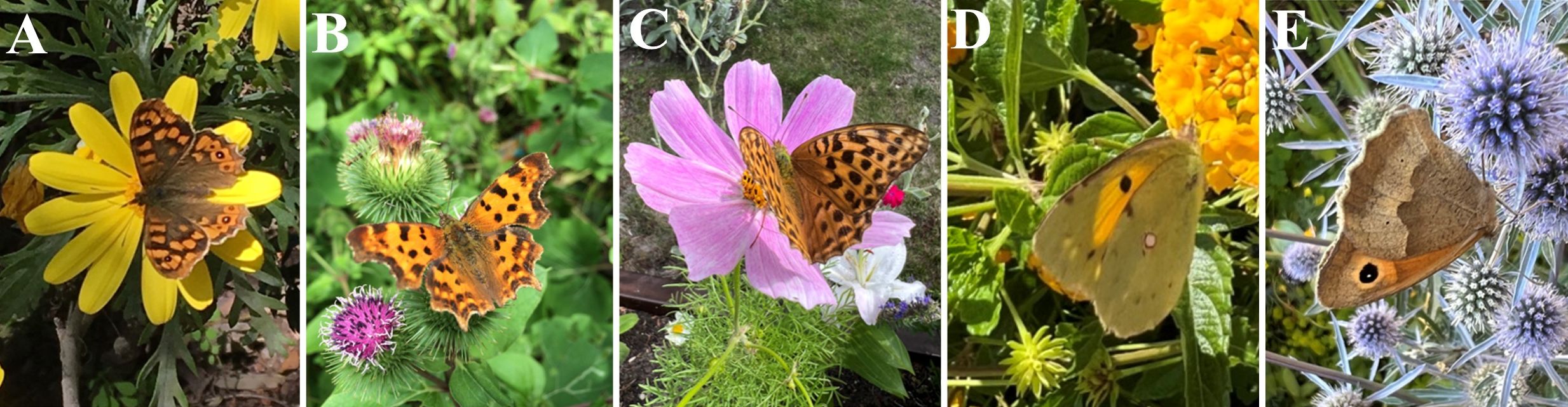}}
    \end{figure}

    \begin{figure}[!ht]
    \setkeys{Gin}{width=1\linewidth}
    \captionsetup[subfigure]{skip=0.5ex,
                             belowskip=1ex,
                             labelformat=simple}
    \renewcommand\thesubfigure{}
    \newcommand\sep{\hspace{0.025\linewidth}}

    \subfloat[Figure 7: Machine learning-supported recognition will be an indispensable tool for recognizing individual morphological features of migratory butterfly specimens in the experimental flower field. It will be crucial to collect data on when and how often the same individual visits the field. Features related to coloration, wing condition, and abdomen size will help determine the generation of the butterflies (individuals arriving from the south typically have lighter wings, which may have serrated edges, and have smaller/narrower abdomens due to the depletion of stored fat resources). A) \textit{Vanessa cardui}, an individual from the migrating spring generation (arriving from the south – photographed in May); B) \textit{V. cardui}, an individual from the autumn generation (photographed in September); C) and D) \textit{Vanessa atalanta}, both individuals from the same summer generation, with slightly different wing patterns, allowing them to be distinguished by the mentioned tool. iPhone 13 Pro photos by Piotr Bębas.]{\includegraphics{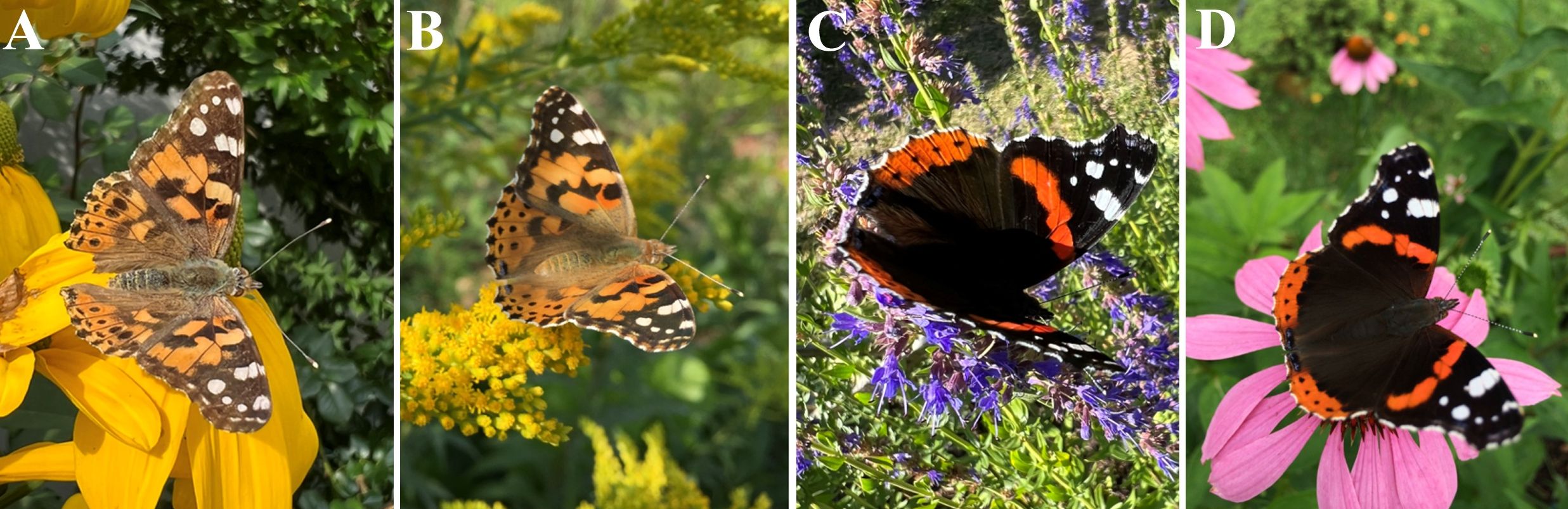}}

    \end{figure}

    \begin{figure}[!ht]
    \setkeys{Gin}{width=1\linewidth}
    \captionsetup[subfigure]{skip=0.5ex,
                             belowskip=1ex,
                             labelformat=simple}
    \renewcommand\thesubfigure{}
    \newcommand\sep{\hspace{0.025\linewidth}}

    \subfloat[Figure 8: Using machine learning-supported recognition for studying territorial behaviors of insects, such as hoverflies, by recognizing individuals and determining their gender, could be groundbreaking and drive the development of zoopsychological research in invertebrates. A) \textit{Eristalis tenax} individual with distinctly darker hind abdominal segments; B) \textit{E. tenax} with abdominal segments featuring orange bands; C) and D) \textit{E. tenax}, respectively female and male, distinguishable based on head structure – the eyes of females are clearly separated (indicated by an arrow and two purple-blue lines), while in males, the eyes almost touch each other (boundary marked by an arrow and one purple-blue line). Photos captured with the HTC Desire 21 Pro 5G phone camera by Piotr Bębas.]{\includegraphics{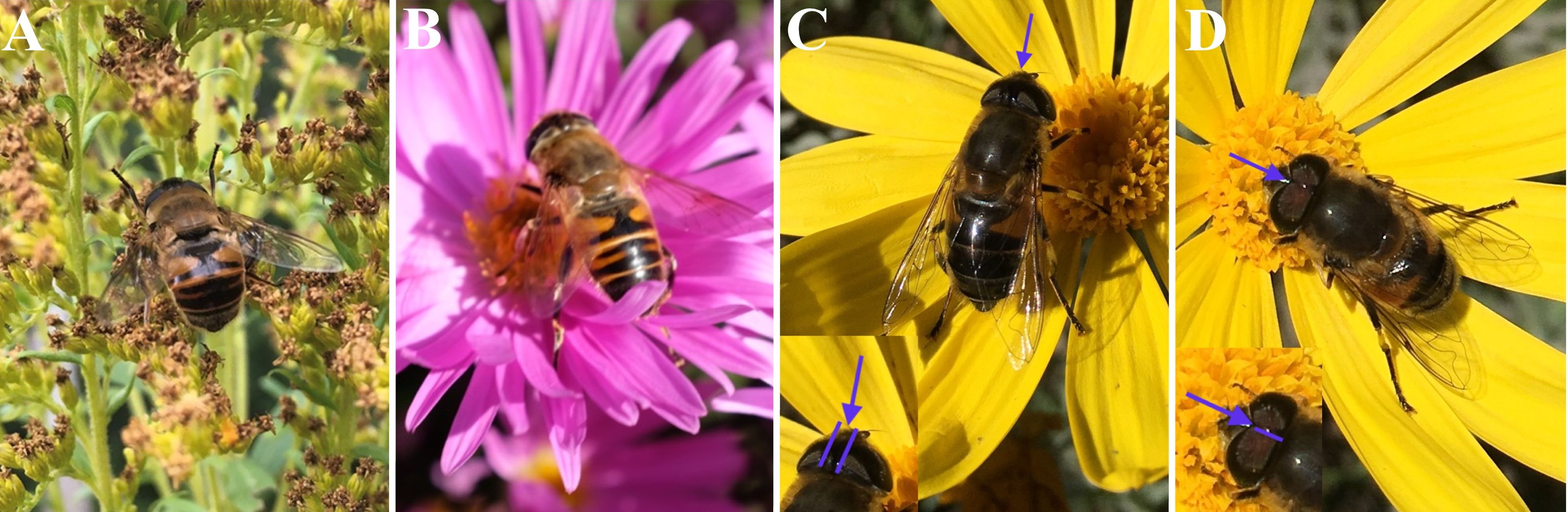}}
    \end{figure}

\section{Panelists}

\paragraph{Izabela Wierzbowska}
Izabela, is a full time academic, working at Jagiellonian University in Kraków, Poland. She holds PhD in biological sciences. Her primary research interests are ecology, environmental sciences and zoology. More specifically, she studies biology and ecology of large mammals, environmental protection, urban ecology, and human-wildlife conflicts. Since 2012 she has been working as an expert for The European Agricultural Fund for Rural Development, and a member of Gorce National Park Scientific Board. She worked in programs evaluated by Polish NGOs within projects on the protection of the Carpathian rivers and mitigation of human – wildlife conflicts in urban areas.

\paragraph{Rafał Stryjek}
Rafał is an associate professor at the Institute of Psychology of the Polish Academy of Sciences in Warsaw. His research centers on the behavior and learning patterns of wild, peri-domestic, and fully-domesticated laboratory rats. His work involves conducting comparative field and laboratory assays to explore differences in neophobia and vigilance 

\paragraph{Piotr Gago}
Piotr Gago is a lecturer at the Polish-Japanese Academy of Information Technology (PJATK), where he engages in research that blends Information and Communication Technologies (ICT) with Social Sciences. He has a keen interest in several areas including software engineering, databases, artificial intelligence, and data science. Piotr also holds a special interest in e-learning, believing in the potential of technology to enhance educational experiences.

\paragraph{Kinga Skorupska}
Assistant professor at Polish-Japanese Academy of Information Technology doing research at the intersection of ICT and Social Sciences. Kinga's interests include online communities and collaboration, user experience, motivation and games. She does research on ICT applications for social good, ranging from inclusion of diverse populations in the main technological discourse, through education to well-being.

\paragraph{Wiesław Kopeć}
Computer scientist, research and innovation team leader, associate professor at Computer Science Faculty of Polish-Japanese Academy of Information Technology (PJAIT). Head of XR Center PJAIT and XR Department. He is also a seasoned project manager with a long-term collaboration track with many universities and academic centers, including University of Warsaw, SWPS University, National Information Processing Institute, and institutes of Polish Academy of Sciences. He co-founded the transdisciplinary HASE research group (Human Aspects in Science and Engineering) and distributed LivingLab Kobo.

\subsection{Research network}
We would like to thank the many people and institutions gathered together by the Living Lab Kobo and HASE Research Group to enable this collaboration and research. First, we would like to thank all the members of HASE research group (Human Aspects in Science and Engineering) and Living Lab Kobo for their support. In particular, the members of XR Center Polish-Japanese Academy of Information Technology (PJAIT) and Emotion-Cognition Lab SWPS University (EC Lab), Kobo Association, in particular, Anna Jaskulska and Maciej Krzywicki, Living Lab Kobo community, especially older adults, for their participation in the lab studies, VR and Psychophysiology Lab of the Institute of Psychology Polish Academy of Sciences as well as National Information Processing Institute (NIPI) for their support in organizing citizen science workshops, especially Jarosław Kowalski. We would like to thank the staff of the Polish National Parks, in particular Jan Loch from Gorce NP and Barbara Pregler from Babia Góra NP. Next, we would like to thank researchers from the Jagiellonian University, especially Izabela Wierzbowska, and her students who participated in workshops on citizen science and machine learning, to co-design a solution that could work for them.

\section{Discussion and Conclusions}

Data scoring is a primary bottleneck in research involving free-ranging animals with motion detection  (see fig. 1,2,3,9,10,11) \cite{STRYJEK2021109303,parsons2023case,parsons2023predator,stryjek2018}. For instance, in studies examining behavioral reactions to predator scents in free-ranging mice (Stryjek et al. unpublished data, and Figure 4), animals were baited for several months to visit chambers where experimental treatments occurred. Three rodent species—striped field mice (\textit{Apodemus agrarius}), yellow-necked mice (\textit{Apodemus flavicollis}), and bank voles (\textit{Clethrionomys glareolus})—(see Fig 4 D-I) visited chambers, with occasional visits from common shrews (\textit{Sorex araneus}). Data on the time and number of entrances, time spent eating, and the number of interactions with a scent probe were collected.

 \begin{figure}[!ht]
    \setkeys{Gin}{width=0.31\linewidth}
    \captionsetup[subfigure]{skip=0.5ex,
                             belowskip=1ex,
                             labelformat=simple}
    \renewcommand\thesubfigure{}
    \newcommand\sep{\hspace{0.025\linewidth}}

    \subfloat[Figure 9: A very clear image of a lynx in the snow, captured by a photo trap located in a Polish national park. This image could be classified using ML as containing an animal.]{\includegraphics{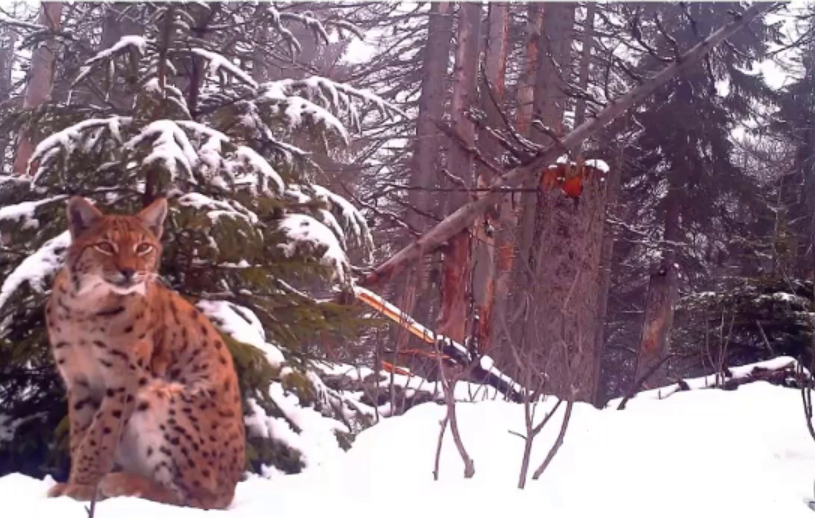}}
    \sep
    \subfloat[Figure 10: An adult male lynx, which, by a researcher working with these animals, can be identified based on its coat pattern.  ]{\includegraphics{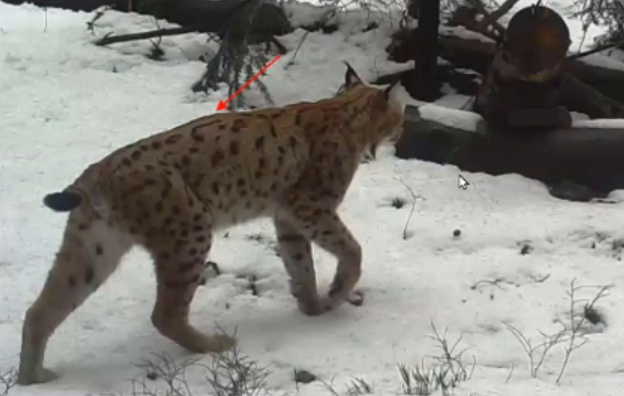}}
    \sep
    \subfloat[Figure 11: The same adult male individual identified due to its unique coat pattern - a task currently impossible for an automated system.]{\includegraphics{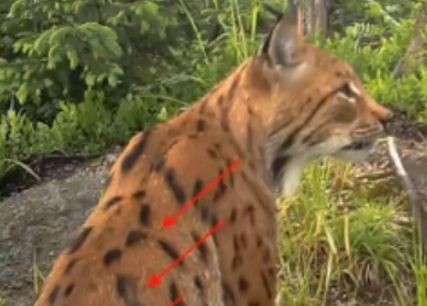}}
    \captionsetup{labelformat=empty}
    \caption{Photos from camera-traps from repository of Gorce NP, used with permission.}
    \end{figure}

Recognizing individuals is also a crucial tool in nearly all rapidly advancing scientific fields of insect science, including ecophysiology, behavioral physiology, and behavioral ecology. A specific focus lies in the ability to identify and distinguish between individuals of the same species e.g. the scarce swallowtail, \textit{Iphiclides podalirius} (Fig. 5 A \& B) and the common brimstone, \textit{Gonepteryx rhamni} (Fig. 5 C \& D) and differentiate morphologically similar species e.g. the speckled wood, \textit{Pararge aegeria} (Fig. 6 A), the comma, \textit{Polygonia c-album} (Fig. 6 B) and the silver-washed fritillary \textit{Argynnis paphia }(Fig. 6 C) and sometimes adopting postures that may hinder identification e.g. the clouded yellow, \textit{Colias croceus} (Fig. 6 D) and the meadow brown, \textit{Maniola jurtina} (Fig. 6 E), particularly in studies on rhythmic behavior (research extended over several days), such as the timing of flower visitation in relation to the functioning of their biological clock mechanism \cite{brady2021,fenske2018}, as well as the biological clock of plants that controls nectar and scent production\cite{fenske2016}, and the reciprocal interactions occurring between plants and insects at the physiological level \cite{fenske2018,bloch2017}. This sets the stage for investigating the coevolution of molecular mechanisms of oscillators and the processes they regulate across different organisms\cite{andersson2002}. A monitoring and identification system for specific individuals proves extremely valuable in field studies of migrating butterflies\cite{stefanescu2016,brattstrom2018}, where the majority of species remain overshadowed by the monarch butterfly (\textit{Danaus plexippus}) and are poorly understood in this context. The role of their oscillators in controlling migration remains entirely unknown – as exemplified by the painted lady, \textit{Vanessa cardui} (Fig. 7 A \& B) and the red admirable, \textit{Vanessa atalanta }(Fig. 7 C \& D). A vital monitoring and identification system is indispensable for investigating insect territorial behavior and patrolling of their occupied territory, such as in males of the common drone fly, \textit{Eristalis tenax} (Fig. 8 A, B, C \& D)\cite{wellington1981}, potentially assessing the role of biological oscillators in this process.

The technical problems that must be addressed in installations of this type are the power supply of data collecting devices, the process of data acquisition (with minimal human involvement, also to avoid influencing the research result), and finally, safe storage of large amounts of collected data.

Manual scoring of recordings for such a study is extremely laborious and usually takes up to several months (R. Stryjek, personal communication). If machine learning could aid in species identification and time spent in chambers, it would save a significant portion of work. Researchers are hopeful for deeper utilization of AI to perform full video tracking (i.e., speed, time spent in various zones), as well as identify specific behaviors (e.g., feeding, different displays of aggressive behavior, courtship). We could also use new technology to track the migration paths of individual species, the development of young animals, and how species interact both with predators and in the predator-prey relationship. These are important issues in protecting not only species as such, but also their habitat, as part of the ecosystem preservation process.

Machine learning also could detect early signs of a rat or mouse infestation on security camera at households and institutions. When we observe damages, feces, or detect a specific scent, it indicates the presence of a considerable group or even a colony of rodents. It is challenging to spot a mouse or rat passing by on a typical surveillance video as they are simply too small to trigger motion detection. However, their movement isn't random; it is typically repetitive, linear, and often along walls. Thus, machine learning can potentially recognize species specific movement patterns, thereby informing users of a possible rodent infestation.

A barrier to the engagement of citizen scientists who are volunteers is that they are not enticed by the idea of spending time doing low-skilled labor either, if it does not provide other benefits. Crowdsourcing tasks need to present some sort of a challenge and appeal to contributors' interests (eg. watching animal behaviors in the wild, or specific interest in bees or the lynx), but they also need to see them as important \cite{Skorupska_Zooniverse}. Tasks such as deciding whether an animal appears in the picture are, in general, not attractive to contributors and their importance is questionable, as "even a kid could do them". Based on our research with older adults, tasks that pose a challenge, such as identifying animal behaviors in a video, or even, identifying individual specimen (Figures 10 and 11), are seen as more skillful and engaging. To ensure sustainable involvement crowd workers' intrinsic motivation ought to be strengthened, either by giving them easy access to unique educational materials (such as mini-documentaries recorded by researchers\footnote{During our citizen science workshops with older adults we have shown a series of video recordings depicting photo trap field work in a Lynx conservation project filmed by J.Loch to see whether they would constitute an attractive edutainment reward.}) and by clearly informing them how their input contributes to larger goals, such as training automated systems or helping specific research and conversation efforts.

Photos classified by the volunteers can then return to the model in a feedback loop, improving its effectiveness, which is important, especially when the initial training set for such rare applications as entity recognition for lynx cases is very limited. Such awareness acts as a driving force for the individuals involved, allowing them to recognize their pivotal role in the process. It enables them to understand that their efforts contribute to making the world a better place, even if one step at a time.

\section*{Acknowledgements}

The Case of Eurasian Lynx research was funded by the Priority Research Area BioS under the program Excellence Initiative – Research University at the Jagiellonian University in Krakow.

\bibliographystyle{splncs04}
\bibliography{bibliography}

\end{document}